\documentclass{ws-ijmpe}
\usepackage[super,compress]{cite}
\begin{document}

\markboth{Xue Pan, Mingmei Xu, Yuanfang Wu}{Instructions for typing manuscripts (paper's title)}

\catchline{}{}{}{}{}

\title{Cumulants and factorial cumulants in the 3-dimensional Ising universality class}

\author{Xue Pan}

\address{School of Electronic Engineering, \\ Chengdu Technological University, Chengdu 611730, China\\
panxue1624@163.com}

\author{Mingmei Xu}

\address{Key Laboratory of Quark and Lepton Physics (MOE) and
Institute of Particle Physics, \\Central China Normal University, Wuhan 430079, China\\}

\author{Yuanfang Wu}

\address{Key Laboratory of Quark and Lepton Physics (MOE) and
Institute of Particle Physics, \\Central China Normal University, Wuhan 430079, China\\
wuyf@mail.ccnu.edu.cn}

\maketitle

\begin{abstract}
The high-order cumulants and factorial cumulants of conserved charges are suggested to study the critical dynamics in heavy ion collisions. In this paper, using parametric representation of the 3-dimensional Ising model, density plots on the phase diagram and temperature dependence of the order parameter cumulants and factorial cumulants are studied and compared. In the vicinity of the critical point, cumulants and factorial cumulants can not be distinguished. Far away from the critical point, sign changes occur in the factorial cumulants comparing with the same order cumulants. The cause of sign changes is analysed. These features may be used to measure the distance to the critical point.
\end{abstract}


\ccode{PACS numbers: 25.75.Nq; 05.50.+q; 64.60.-i; 24.60.-k}


\section{Introduction}

Searching for the critical point and locating the phase boundary of quantum chromo-dynamics (QCD) phase transition are main goals of current relativistic heavy-ion collision experiments~\cite{maingoal}. The high-order cumulants of conserved charges, especially for net-baryon number, have been studied for many years. They reflect the fluctuations of conserved charges. They are more sensitive to the correlation length and can be measured and calculated by experiments and theory, respectively\cite{stephanov-prl91, koch, Stephanov-prl102, Karsch-EPJC71}. Lots of results of the high-order cumulants have been presented from experiments and theory\cite{Asakawa-prl103, Stephanov-prl107, Fuweijie, Vladi, JournalofHighEnergyPhysics2018205, PhysicalReviewD101074502, PhysRevLett112032302, PhysRevLett113092301, PhysLettB785551-560}. The sign changes and non-monotonic behavior of the high-order cumulants are considered to be related to the critical signals.

Recently, the other kind of moments, the factorial cumulants, which are also known as the integrated multi-particle correlations, get a lot of attention~\cite{PhysLettB728386-392, NuclearPhysicsA94265-96, PhysRevC93034915, PhysRevC95064912, PhysRevC95054906, PhysLettB774623-629, PhysRevC96024910, EurPhysJC77288, PhysRevC100051902}. On one hand, they have advantages in the analysis of the acceptance dependence in experiments~\cite{PhysRevC93034915}. They are useful in the reconstruction of the original cumulants from the incomplete information obtained experimentally~\cite{PhysRevC86044904, PhysRevC96024910}. The uses of the factorial cumulants in the analysis of data in the heavy ion collisions have been proposed~\cite{PhysRevC95054906, PhysRevC100051902, EurPhysJC77288}. On the other hand, multi-proton correlations have been found in the STAR data, at least at the lower energies~\cite{PhysRevC95054906, EurPhysJC77288}. In Ref.~\refcite{PhysRevC95054906}, using the parametric representation of Ising model, the authors showed that the signs of the second to fourth order factorial cumulants are a useful tool to exclude regions in the QCD phase diagram close to the critical point. In the other model, very large values for the fifth and sixth order factorial cumulants have been predicted, which can be tested in experiments~\cite{PhysRevC98054901}.

The high-order cumulants are related to the generalized susceptibilities which can be got from the derivatives of the free energy with respect to the chemical potential. Their properties are explicable with that clear definition. While the $n$-th order factorial cumulant of particle distribution reflects the integrated $n$-particle correlation, which can not be calculated directly. One way is to express and calculate the factorial cumulants indirectly by the cumulants~\cite{PhysRevC95054906, PhysRevC96024910}. In fact, cumulants and factorial cumulants can not be distinguished and can be seen as identical in the vicinity of the critical point in a model of critical fluctuations as discussed in Ref.~\refcite{PhysRevC93034915}.

To make clear the meaning of the signs of factorial cumulants, and the relation to cumulants, further study and comparison should be useful.

The QCD critical point, if exists, is expected to belong to the 3-dimensional Ising universality class~\cite{class1, class2, class3, class4}. Critical behavior of the thermodynamics, which is controlled by the critical exponents, is the same in the same universality class. Recently, many works have been done to map the results of the 3-dimensional Ising model to that of the QCD~\cite{PhysRevD102014505, PhysRevC103034901}. Usually, a linear correspondence between the QCD variables, temperature and net-baryon chemical potential, and the Ising variables, temperature and magnetic fiels are supposed~\cite{linearmap1, linearmap2, linearmap3, NPA}. Cumulants of the net-baryon number, which are the derivatives of the QCD free energy density with respect to net-baryon chemical potential, can be regarded as the combination of the derivatives with respect to temperature and magnetic field in the 3-dimensional Ising model in the vicinity of the critical point. Since the critial exponent of magnetic field is larger than that of the temperature~\cite{Ising exponents1}, the critical behavior of net-baryon number fluctuations is mainly controlled by the derivatives with respect to the magnetic field, i.e., the fluctuations of the order parameter in the 3-dimensional Ising model.

By using the parametric representation of the 3-dimensional Ising model, we study the signs of the third to sixth order cumulants and factorial cumulants of the order parameter on the phase diagram. The reason of sign changes of factorial cumulants relative to cumulants is analysed. Temperature dependence of the second to sixth order factorial cumulants is discussed and compared to that of the cumulants at different distances to the phase boundary.

The paper is organized as following. In Section 2, the Ising model in the parametric representation is introduced. The parametric expressions of the second to sixth order cumulants are derived. The relation between the factorial cumulants and cumulants is presented. In Section 3, density plots of the third to sixth order factorial cumulants and cumulants on the phase diagram are studied and compared. In Section 4, temperature dependence of the second to sixth order cumulants and factorial cumulants is shown and discussed. Finally, the conclusions and summary are given in Section 5.

\section{Cumulants and factorial cumulants in the parametric representation}

There are two parameters in the 3-dimensional Ising model, the temperature $T$ and the external magnetic field $H$. Using $T_c$ representing the critical temperature, in the parametric representation, the magnetization $M$ (order parameter of the Ising model) and reduced temperature $t=(T-T_c)/{T_c}$ can be parameterized by two variables $R$ and $\theta$~\cite{linearpara, linearpara3},
\begin{equation}\label{parametric}
M=R^{\beta}\theta,~~~~~~t=R(1-\theta^2).
\end{equation}

The equation of state of the 3-dimensional Ising model can be given by the parametric representation in terms of $R$ and $\theta$ as
\begin{equation}\label{equation state}
 H=R^{\beta\delta}h(\theta).
\end{equation}
Where $\beta$ and $\delta$ are critical exponents of the three-dimensional Ising universality class with values 0.3267(10) and 4.786(14), respectively~\cite{Isingexponents}. Because $M$ is an odd function of the magnetic field $M(-H)=-M(H)$, the function $h(\theta)$ should be an odd function of $\theta$. What is more, $h(\theta)$ is analytic. One simple choice of $h(\theta)$ in the form of the linear parametric model is as follows~\cite{linearpara},
\begin{equation}\label{equation h}
h(\theta)=\theta(3-2\theta^{2}).
\end{equation}

The $n$-th order cumulant of order parameter in the Ising model can be got from the derivatives of $M$ with respect to $H$,
\begin{equation}\label{cumulants}
 \left.\kappa_{n}(t,H)=(\frac{\partial^{n-1} M}{\partial H^{n-1}})\right|_{t}.
\end{equation}
When taking the approximate values of the critical exponents $\beta=1/3$ and $\delta=5$, the first six order cumulants in the parametric presentations are as follows,

\begin{equation}\label{first-order cumulants}
\kappa_{1}(t,H)=R^{1/3}\theta
\end{equation}

\begin{equation}\label{second-order cumulants}
\kappa_{2}(t,H)=\frac{1}{R^{4/3}(2\theta^2+3)},
\end{equation}

\begin{equation}\label{third-order cumulants}
\kappa_{3}(t,H)=\frac{4\theta(\theta^2+9)}{R^{3}(\theta^2-3)(2\theta^2+3)^3},
\end{equation}

\begin{equation}\label{fourth-order cumulants}
\kappa_{4}(t,H)=12\frac{(2\theta^8-5\theta^6+105\theta^4-783\theta^2+81)}{R^{14/3}(\theta^2-3)^3(2\theta^2+3)^5},\\
\end{equation}

\begin{equation}\label{fifth-order cumulants}
\kappa_{5}(t,H)=48\theta\frac{\sum_{n=0}^{n=6}{a_{2n}\theta^{2n}}}{R^{19/3}(\theta^2-3)^5(2\theta^2+3)^7},\\
\end{equation}

\begin{equation}\label{sixth-order cumulants}
\kappa_{6}(t,H)
=240\frac{\sum_{n=0}^{n=9}{b_{2n}\theta^{2n}}}{R^8(\theta^2-3)^7(2\theta^2+3)^9},
\end{equation}
where $a_{2n}$ and $b_{2n}$ are the expansion coefficients of the fifth and sixth order cumulants. Their values are listed in Table 1 and Table 2, respectively.

\begin{table}[pt]
\tbl{ Values for the expansion coefficients $a_{2n}$.}
{\begin{tabular}{@{}ccccccc@{}} \toprule
$a_0$ ~&$a_2$~ &$a_4$ ~&$a_6$ ~& $a_8$ ~~ &$a_{10}$ ~~ & $a_{12}$\\
-18225~ & 86670 ~& -33966 ~ & 8244 ~& -741 ~~& 14 ~~& 4\\ \botrule
\end{tabular}}
\end{table}

\begin{table}[pt]
\tbl{Values for the expansion coefficients $b_{2n}$.}
{\begin{tabular}{@{}cccccccccc@{}} \toprule
$b_0$ ~&$b_2$ ~&$b_4$ ~&$b_6$ ~& $b_8$ ~&$b_{10}$ ~~ & $b_{12}$ ~~ & $b_{14}$ ~~& $b_{16}$ ~~ & $b_{18}$ \\
-98415~ & 3306744~ & -11234619~ & 7120872~ &-2736261~~ & ~~501120~~ &-53001 ~~& 1560 ~~ & -8 ~~& 8 \\ \botrule
\end{tabular}}
\end{table}

The first six order factorial cumulants ($C_n$) can be expressed by the cumulants as follows~\cite{PhysRevC96024910},
\begin{equation}\label{first-order factorial cumulants}
C_{1}=\kappa_{1},
\end{equation}
\begin{equation}\label{second-order factorial cumulants}
C_{2}=\kappa_{2}-\kappa_{1},
\end{equation}
\begin{equation}\label{third-order factorial cumulants}
C_{3}=\kappa_{3}-3\kappa_{2}+2\kappa_{1},
\end{equation}
\begin{equation}\label{fourth-order factorial cumulants}
C_{4}=\kappa_{4}-6\kappa_{3}+11\kappa_{2}-6\kappa_{1},
\end{equation}
\begin{equation}\label{fifth-order factorial cumulants}
C_{5}=\kappa_{5}-10\kappa_{4}+35\kappa_{3}-50\kappa_{2}+24\kappa_{1},
\end{equation}
\begin{equation}\label{sixth-order factorial cumulants}
C_{6}=\kappa_{6}-15\kappa_{5}+85\kappa_{4}-225\kappa_{3}+274\kappa_{2}-120\kappa_{1}.
\end{equation}

From Eq.~\eqref{first-order cumulants} to Eq.~\eqref{sixth-order cumulants}, it is clear that the odd-order cumulants are odd functions of $\theta$, while the even-order cumulants are even functions of $\theta$. Because $H$ is also an odd function of $\theta$, so the odd-order and even-order cumulants are odd and even functions of $H$, respectively. The relation can be expressed as follows,
\begin{equation}\label{odd or even of cumulants}
\kappa_{2n-1}(-H)=-\kappa_{2n-1}(H), ~~\kappa_{2n}(-H)=\kappa_{2n}(H),~~n=1,2,3....
\end{equation}
In Ref.~\refcite{Stephanov-prl107} and Ref.~\refcite{cpc-ising-six}, it has shown that density plots of the fourth and sixth order cumulants on the phase diagram are symmetric around the axis $H=0$.

Turn to the high-order factorial cumulants, because they mix each order of cumulants, they are not odd or even functions of the magnetic field any more. Density plots of the second to the fourth order factorial cumulants do not have symmetries about the magnetic field any more as showed in Ref.~\refcite{PhysRevC95054906}. What is more, because the higher the order of the cumulants, the more sensitive to the correlation length, one can predict from Eq.~\eqref{first-order factorial cumulants} to Eq.~\eqref{sixth-order factorial cumulants} that the highest cumulant will dominate the behavior of the corresponding factorial cumulant in the vicinity of the critical point. When far away from the critical point, this dominance will disappear.

\section{Density plots of the third to sixth order cumulants and factorial cumulants}

To compare the cumulants and factorial cumulants clearly, using the parametric representation of the 3-dimensional Ising universality class, density plots of $\kappa_3$ to $\kappa_6$, $C_3$ to $C_6$ on the $(H,t)$ plane are studied and shown in Fig.~1(a)-(h), respectively. The critical point is indicated by a red dot, the first-order phase transition line is represented by the solid black line, while the crossover is presented by the dashed black line, respectively. The white areas correspond to positive values of $\kappa_n$ and $C_n$, while the green ones correspond to negative values.

\begin{figure}[hbt]
\centering
    \includegraphics[width=0.254\textwidth]{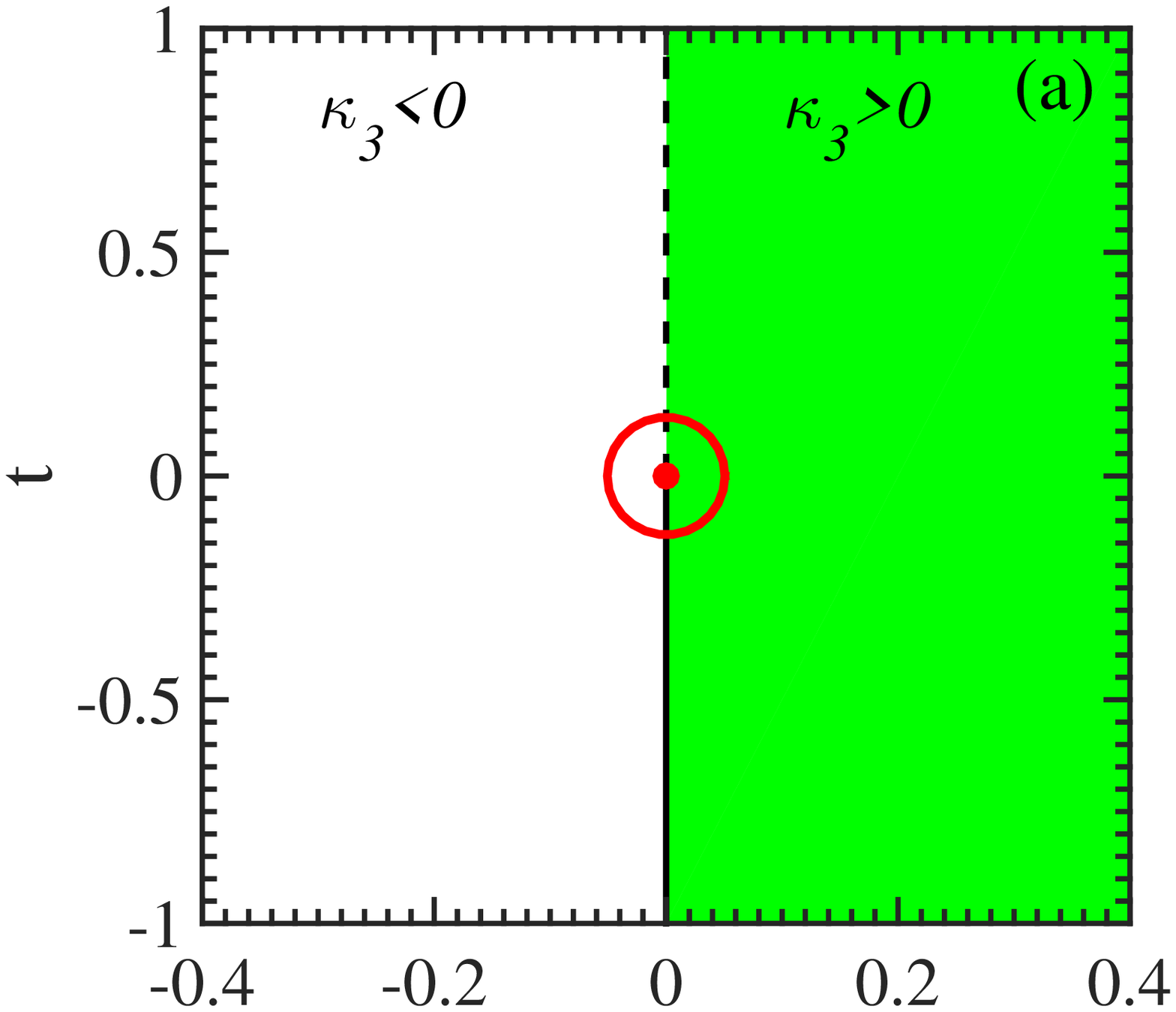}
    \includegraphics[width=0.239\textwidth]{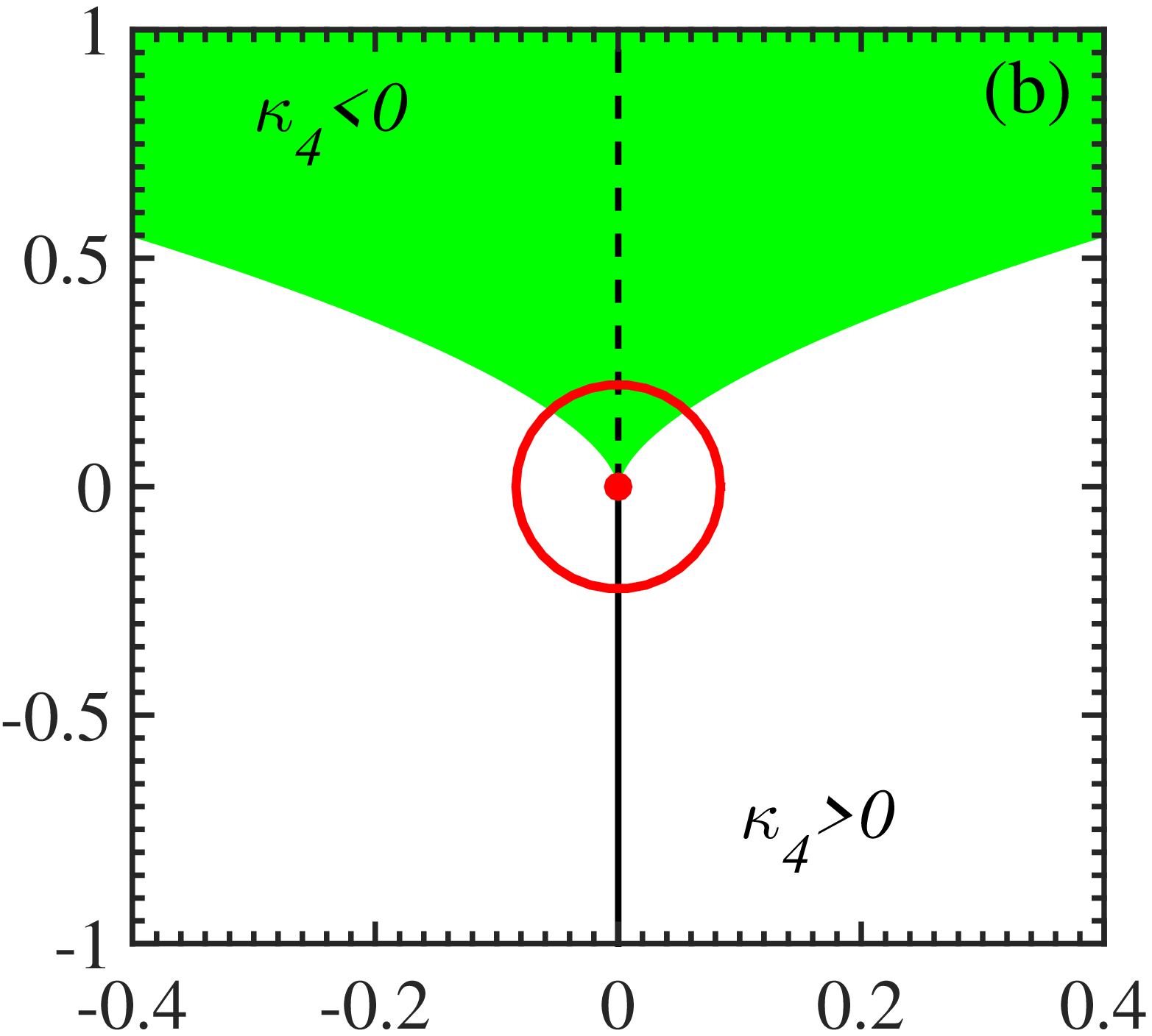}
    \includegraphics[width=0.239\textwidth]{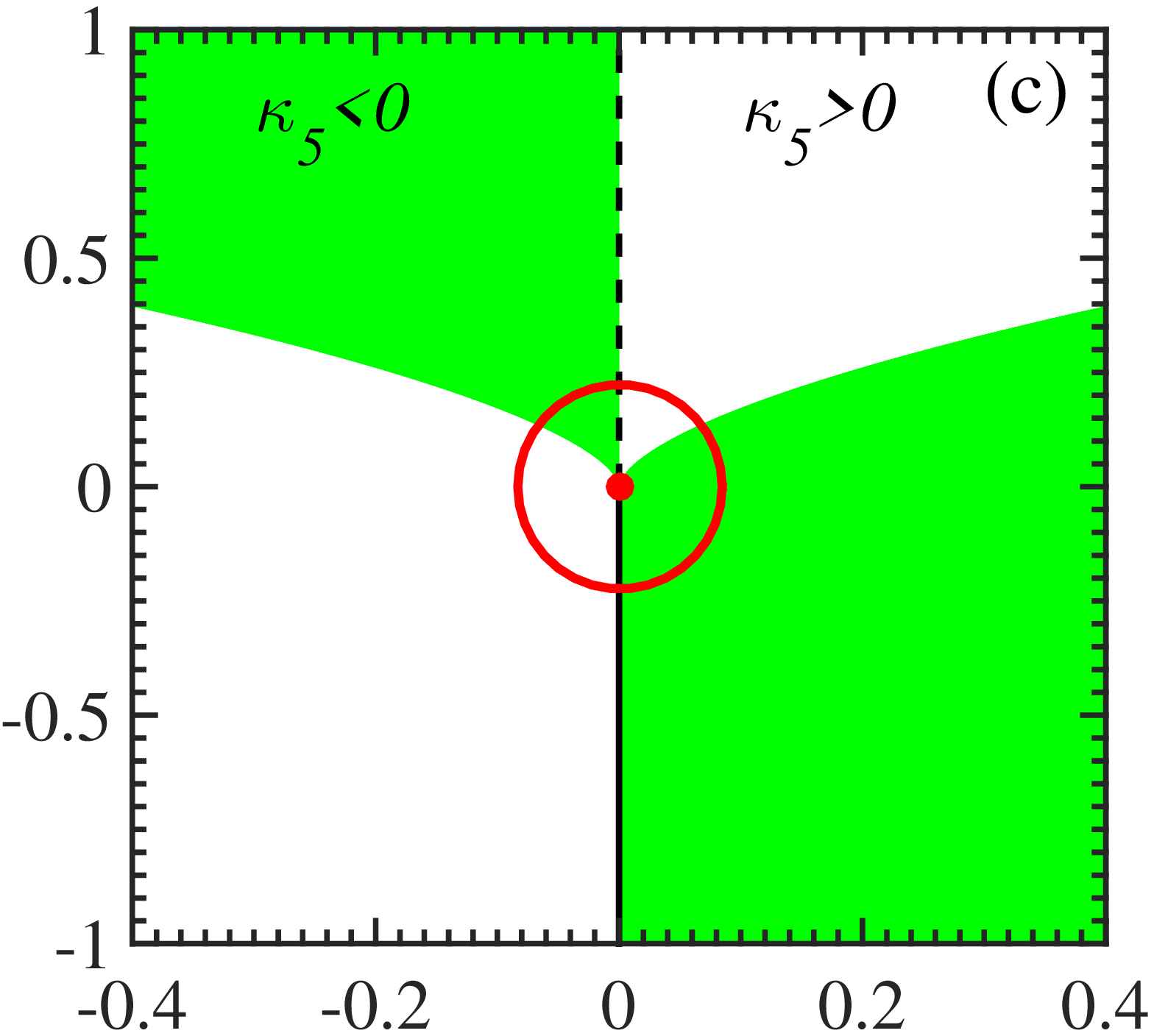}
    \includegraphics[width=0.239\textwidth]{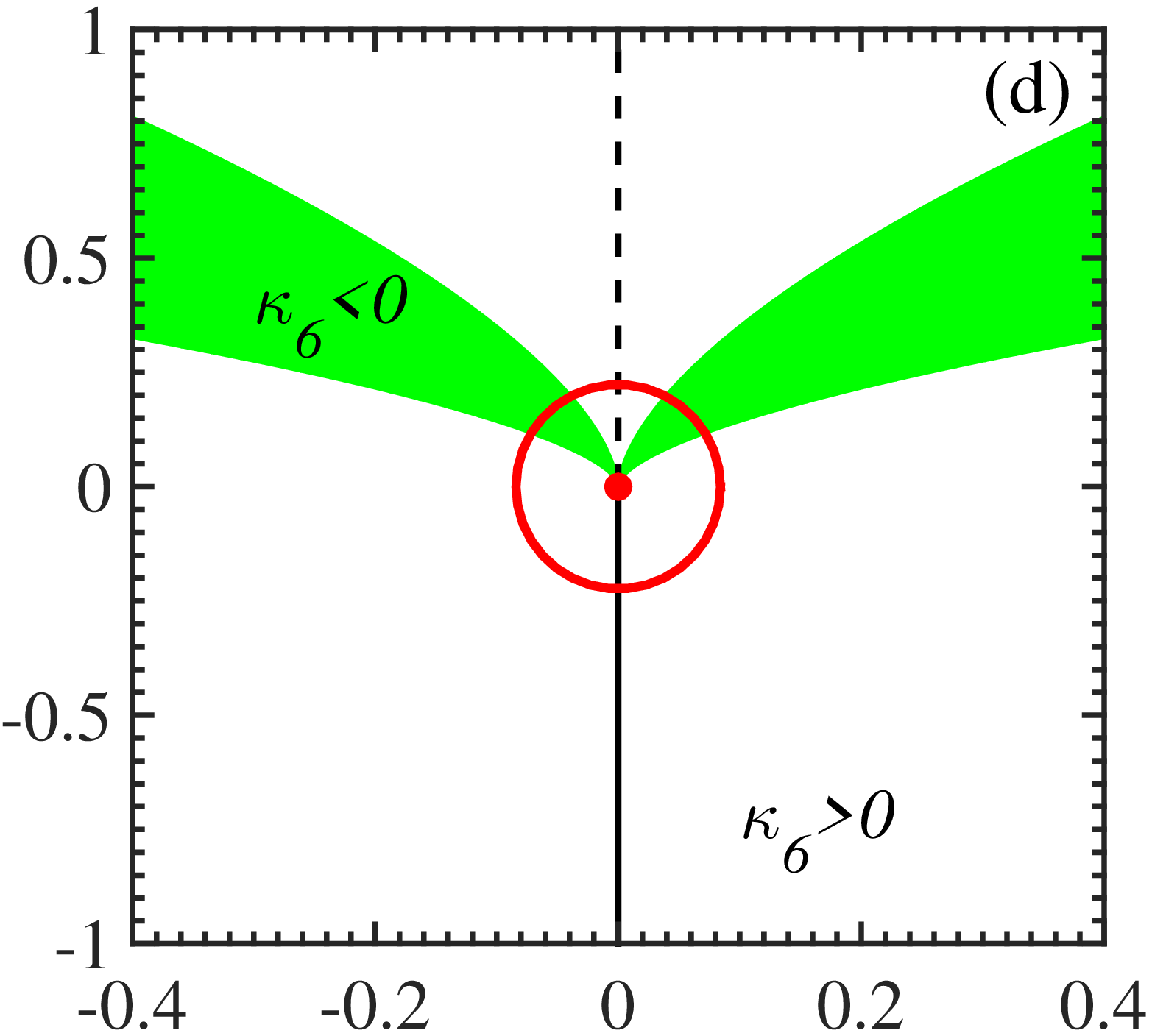}
    \includegraphics[width=0.257\textwidth]{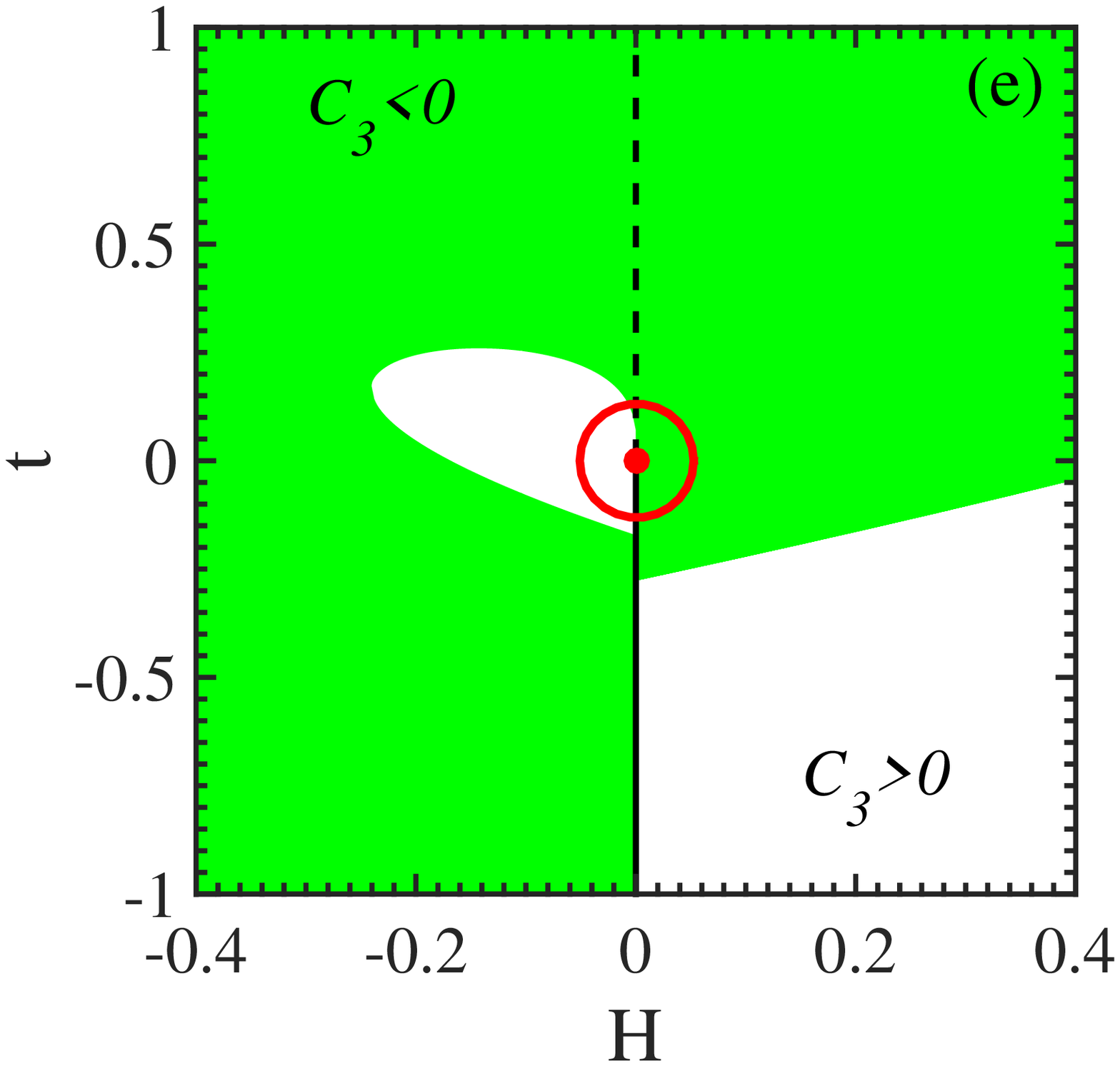}
    \includegraphics[width=0.239\textwidth]{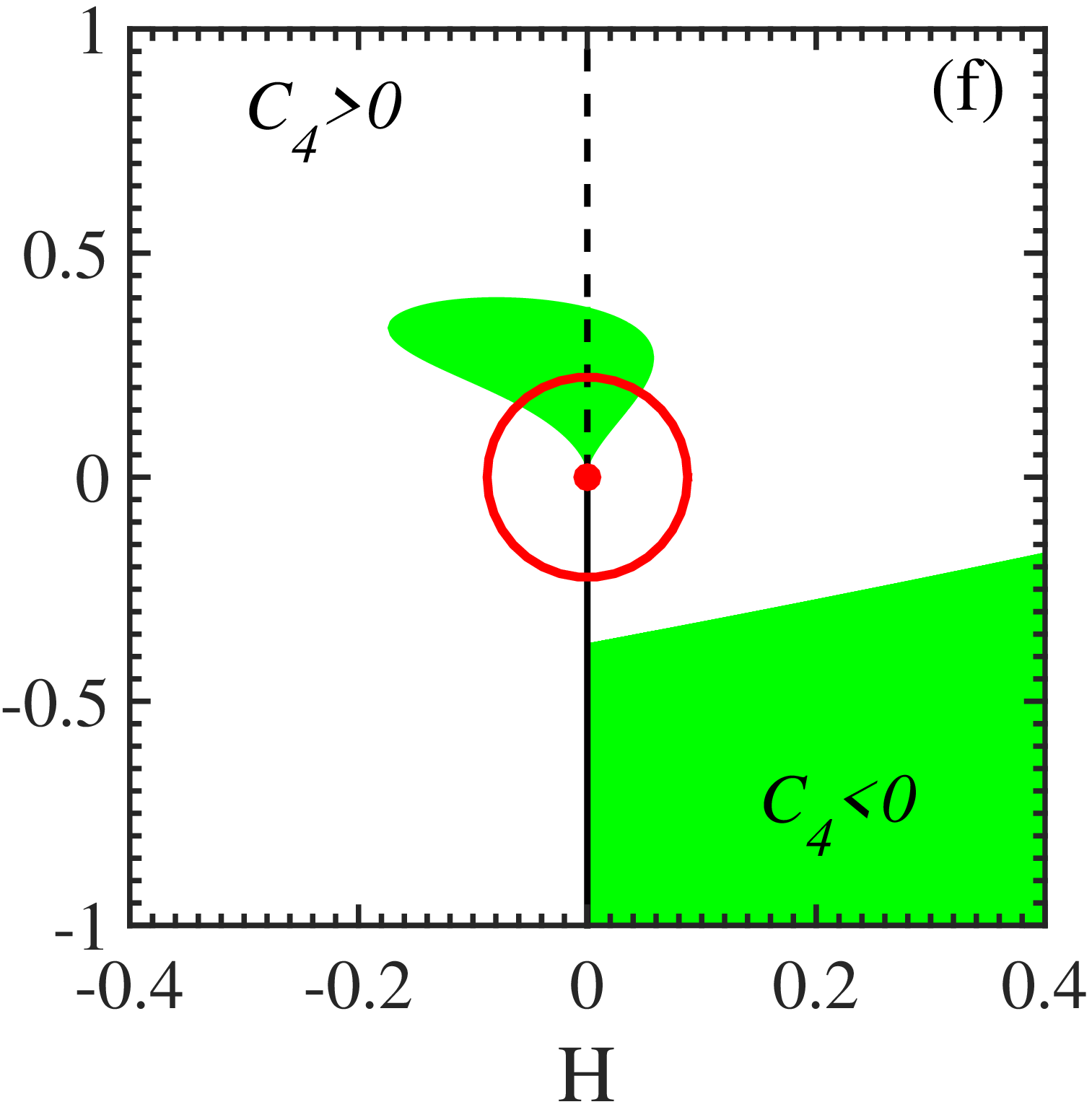}
    \includegraphics[width=0.239\textwidth]{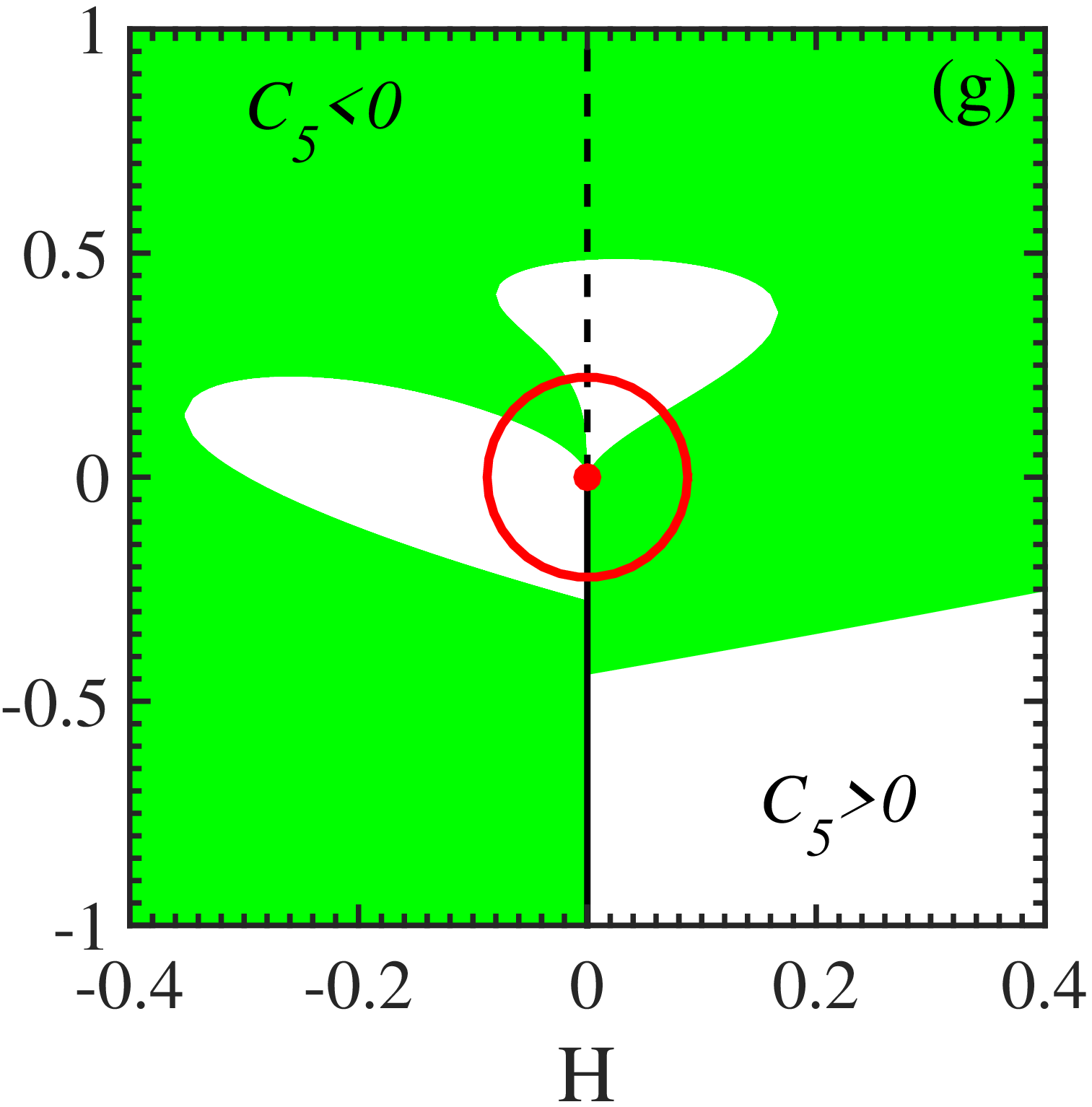}
    \includegraphics[width=0.239\textwidth]{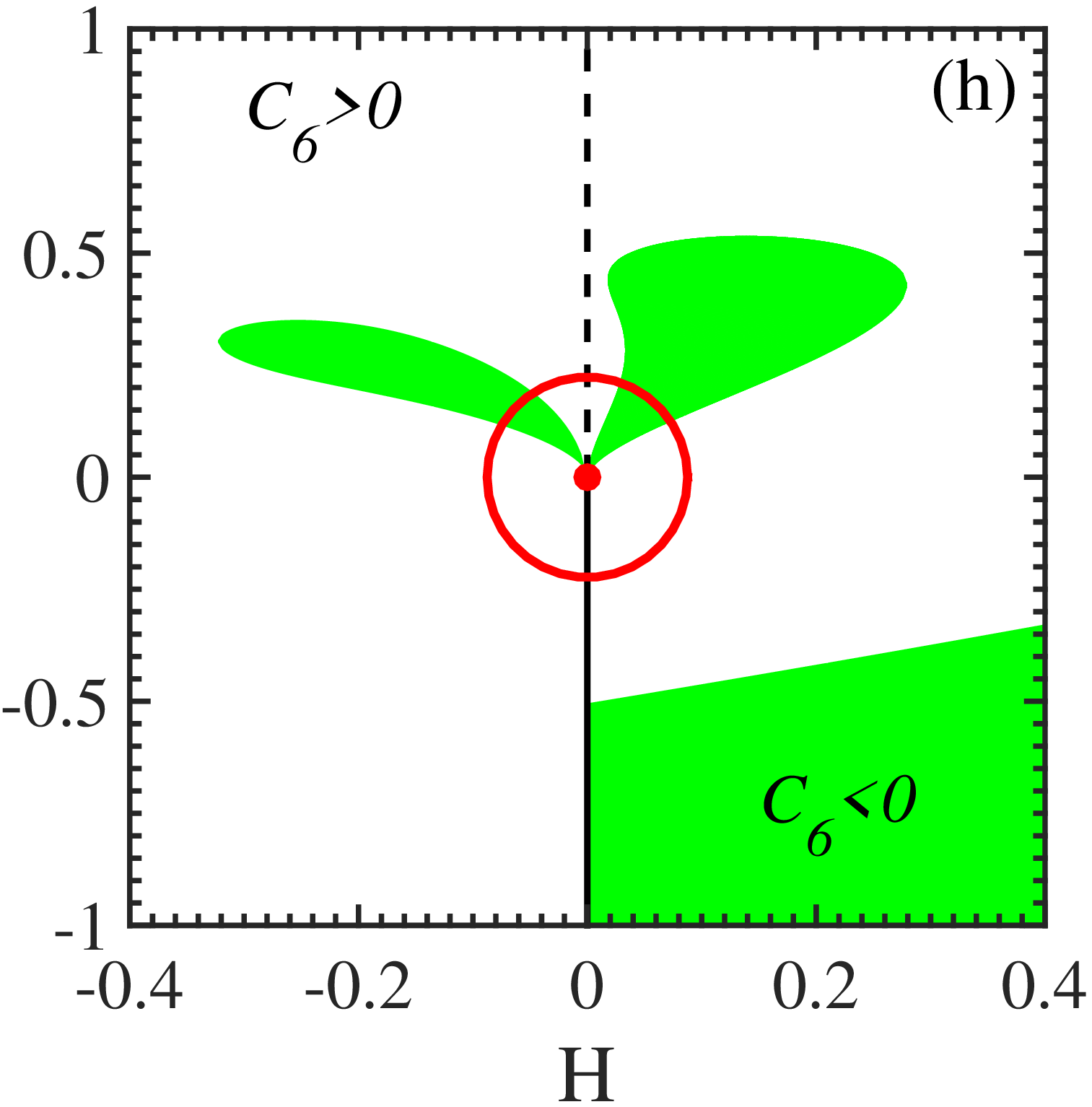}
    \caption{\label{figure 1}(Color online). Density plots of $\kappa_3$ to $\kappa_6$ (upper panels) and $C_3$ to $C_6$ (lower panels) in the $(H,t)$ plane. The
     critical point is indicated by a red dot, the first-order phase transition line is represented by the solid black line, while the crossover is presented by the dashed black line, respectively. The white areas correspond to positive values of $\kappa_n$ and $C_n$, while the green ones correspond to negative values. The red circles show an area where density plots of $\kappa_n$ and $C_n$ are similar.}
\end{figure}

Density plots of $\kappa_3$ to $\kappa_6$ conform to the parity in Eq.~\eqref{odd or even of cumulants}. Signs of $\kappa_3$ are opposite at the two sides of axis $H=0$, so are those of $\kappa_5$ as shown in Fig.~1(a) and (c), respectively. While signs are the same for $\kappa_4$ and $\kappa_6$ as shown in Fig.~1(b) and (d), respectively. In Fig.~1(e) to (h), signs of $C_3$ to $C_6$ have no symmetries about the axis $H=0$.

Comparing the up panels and the lower panels in Fig. 1, it is clear that density plots of the same order cumulant and factorial cumulant are similar in the vicinity of the critical point, such as in the area surrounded by red circles. This is because that the behavior of $C_n$ is dominated by $\kappa_n$ in the vicinity of the critical point.

When far away from the critical point, the similarity of $C_n$ and $\kappa_n$ disappears. For example, at the lower temperature side and positive magnetic field, signs of $C_n$ and $\kappa_n$ are opposite as shown in the lower right corners of Fig. 1(a) and (e), (b) and (f), (c) and (g), (d) and (h) for $n=3, 4, 5, 6$, respectively. In fact, in the lower temperature side, spins of the Ising model are arranged in order, the system is in ordered phase. As decreasing of temperature, the other high-order cumulants approaching zero, while the first order one, the magnetization, becomes bigger and gets saturation. So signs of $C_n$ are dominated by the term related to $\kappa_1$ in Eq.~\eqref{first-order factorial cumulants} to Eq.~\eqref{sixth-order factorial cumulants} at the lower temperature side.

Take $C_6$ as an example and show it in detail. $C_6$ is made up of six terms in Eq.~\eqref{sixth-order factorial cumulants}. At the lower temperature side and a positive magnetic field, signs of $C_6$ and the six terms are shown in Table 3.

\begin{table}[pt]
\tbl{Sign of $C_6$ and each term at the lower temperature side and positive magnetic field.}
{\begin{tabular}{@{}ccccccc@{}} \toprule
$C_6$~~~~ & $\kappa_6$ ~~~~& -15$\kappa_5$ ~~~~& 85$\kappa_4$ ~~~~&-225$\kappa_3$ ~~~~& 274$\kappa_2$ ~~~~& -120$\kappa_1$\\
-~~~~ & + ~~~~& + ~~~~& + ~~~~& + ~~~~& + ~~~~& - ~\\ \botrule
\end{tabular}}
\end{table}

It is clear that the sign of $C_6$ is opposite with the first five terms, while it is the same with the last term which is related to $\kappa_1$. As we know, the lower the temperature, the more ordered the system, the bigger the value of magnetization. When approaching to the lower temperature side, the absolute value of $\kappa_1$ (the magnetization) gets bigger while the higher order cumulants are approaching to zero. The last term in Eq.~\eqref{first-order factorial cumulants} to Eq.~\eqref{sixth-order factorial cumulants} determines that the even order factorial cumulants have different sign with $\kappa_1$ and the odd order have the same sign with $\kappa_1$. What is more, $\kappa_1$ is positive at $H>0$ and negative at $H<0$. That explains why $C_4$ and $C_6$ are negative at $H>0$ and positive at $H<0$, while $C_3$ and $C_5$ are reversed.

Overall, at the lower temperature side far away from the critical temperature, signs of $C_n$ depend on the last term in Eq.~\eqref{first-order factorial cumulants} to Eq.~\eqref{sixth-order factorial cumulants} which is related to $\kappa_1$. Comparing $C_n$ and $\kappa_n$, if their signs are opposite for each order, it may predict that the system is in the ordered phase and far away from the critical point.

At the higher temperature side, spins of the Ising model are disordered. The system is in the disordered phase. As increasing of temperature, $\kappa_1$ or the magnetization is approaching to zero, the last term in Eq.~\eqref{first-order factorial cumulants} to Eq.~\eqref{sixth-order factorial cumulants} will not dominate the sign of $C_n$ any more. The influences of the other terms related to the lower orders of cumulants show up. For example, in the up right corner of Fig. 1(c) and (g), the sign of $\kappa_5$ and $C_5$ is opposite. Signs of $C_5$ and each term made up of $C_5$ in Eq.~\eqref{fifth-order factorial cumulants} are shown in Table 4. It is clear that the sign of $C_5$ is consistent with the terms which are related to $\kappa_2$ and $\kappa_3$. It is opposite with the terms related to higher order cumulants $\kappa_4$ and $\kappa_5$, and also $\kappa_1$.

\begin{table}[pt]
\tbl{Sign of $C_5$ and each term at the higher temperature side and positive magnetic field.}
{\begin{tabular}{@{}cccccc@{}} \toprule
$C_5$~~~~ &$\kappa_5$ ~~~~ & -10$\kappa_4$ ~~~~ &35$\kappa_3$ ~~~~ & -50$\kappa_2$ ~~~~ & 24$\kappa_1$\\
-~~~~ & + ~~~~ & + ~~~~ & - ~~~~ & - ~~~~ & + ~\\ \botrule
\hline
\end{tabular}}
\end{table}

In summary, in the vicinity of the critical point, density plots of $C_n$ and $\kappa_n$ are still consistent. The higher the order of the cumulants, the more sensitive to the correlation length, the stronger of the singularity. The first term of $C_n$ at the right of Eq.~\eqref{first-order factorial cumulants} to Eq.~\eqref{sixth-order factorial cumulants}, i.e. $\kappa_n$, dominates its sign. Far away from the critical point, the terms on the right side of Eq.~\eqref{first-order factorial cumulants} to Eq.~\eqref{sixth-order factorial cumulants} which related to the lower orders of cumulants will influence the sign of $C_n$, resulting the sign difference between $C_n$ and $\kappa_n$. The comparison of signs of $C_n$ and $\kappa_n$ can give some information about the distance to the critical point. To learn more about their behavior, at different distances to the phase boundary, their temperature dependence is studied in the next section.

\section{Temperature dependence of the cumulants and factorial cumulants}

The temperature dependence of the second to sixth order cumulants and factorial cumulants at different positive and negative magnetic fields is studied, respectively. The bigger the absolute value of $H$, the further away from the phase boundary.

Temperature dependence of the even-order cumulants $\kappa_{2n}$ and factorial cumulants $C_{2n}$ for $n=1, 2, 3$ at magnetic fields $H=0.01, 0.05, 0.2, -0.05$ are shown in Fig. 2(a)-(d), respectively. For the sake of comparison, they have been normalized by their maximum. For $\kappa_{2n}$, the qualitative temperature dependence does not change with magnetic field. $\kappa_2$ keeps a peak structure and positive. At all of the magnetic fields, temperature dependence of $\kappa_4$ and $\kappa_6$ oscillates. The ratios of maximum to minimum for $\kappa_4$ and $\kappa_6$ approximate $-28$ and $-6$, respectively~\cite{cpc-ising-six, Stephanov-prl107}.

\begin{figure*}[hbt]
\centering
    \includegraphics[width=0.4\textwidth]{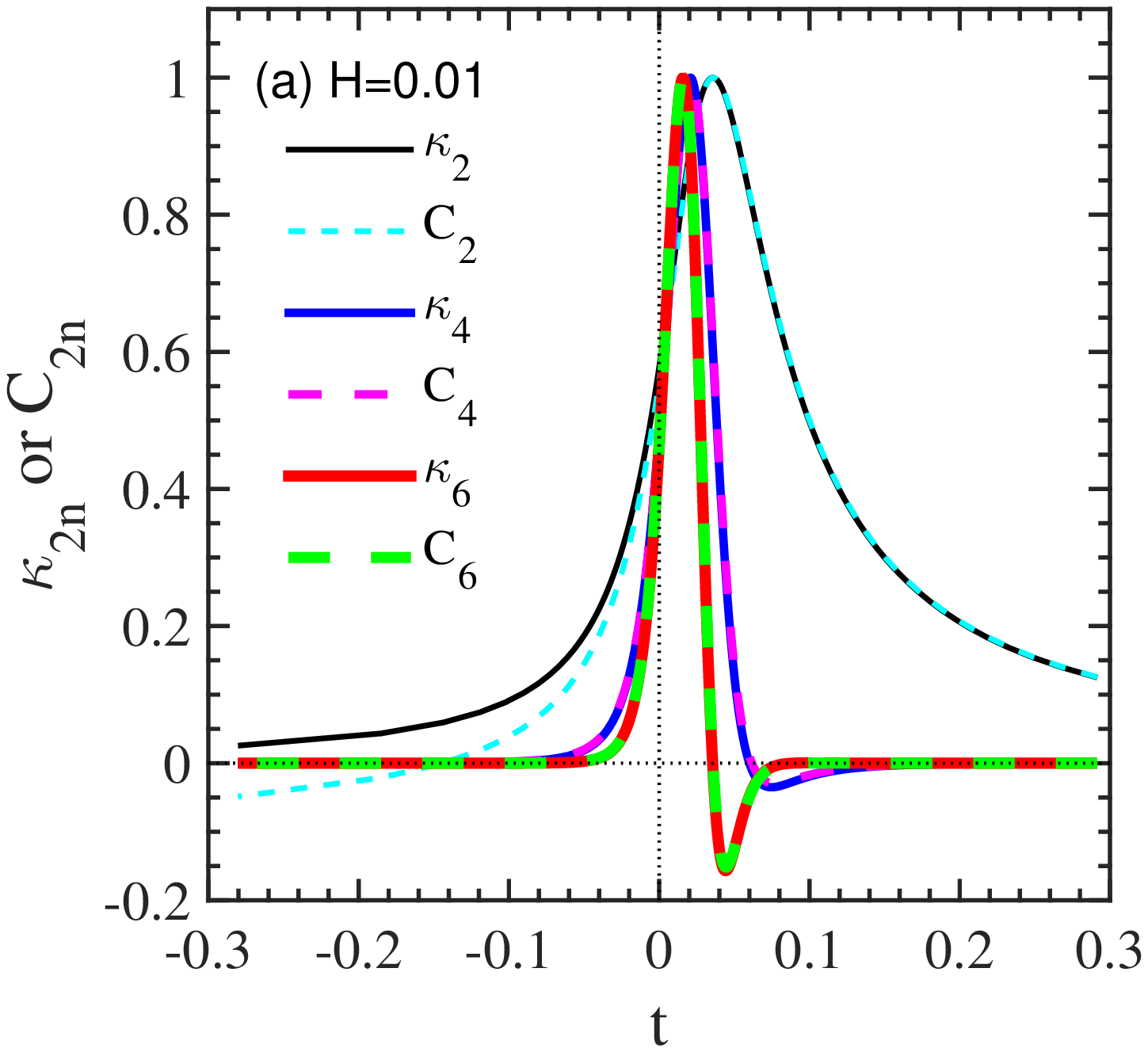}
    \includegraphics[width=0.4\textwidth]{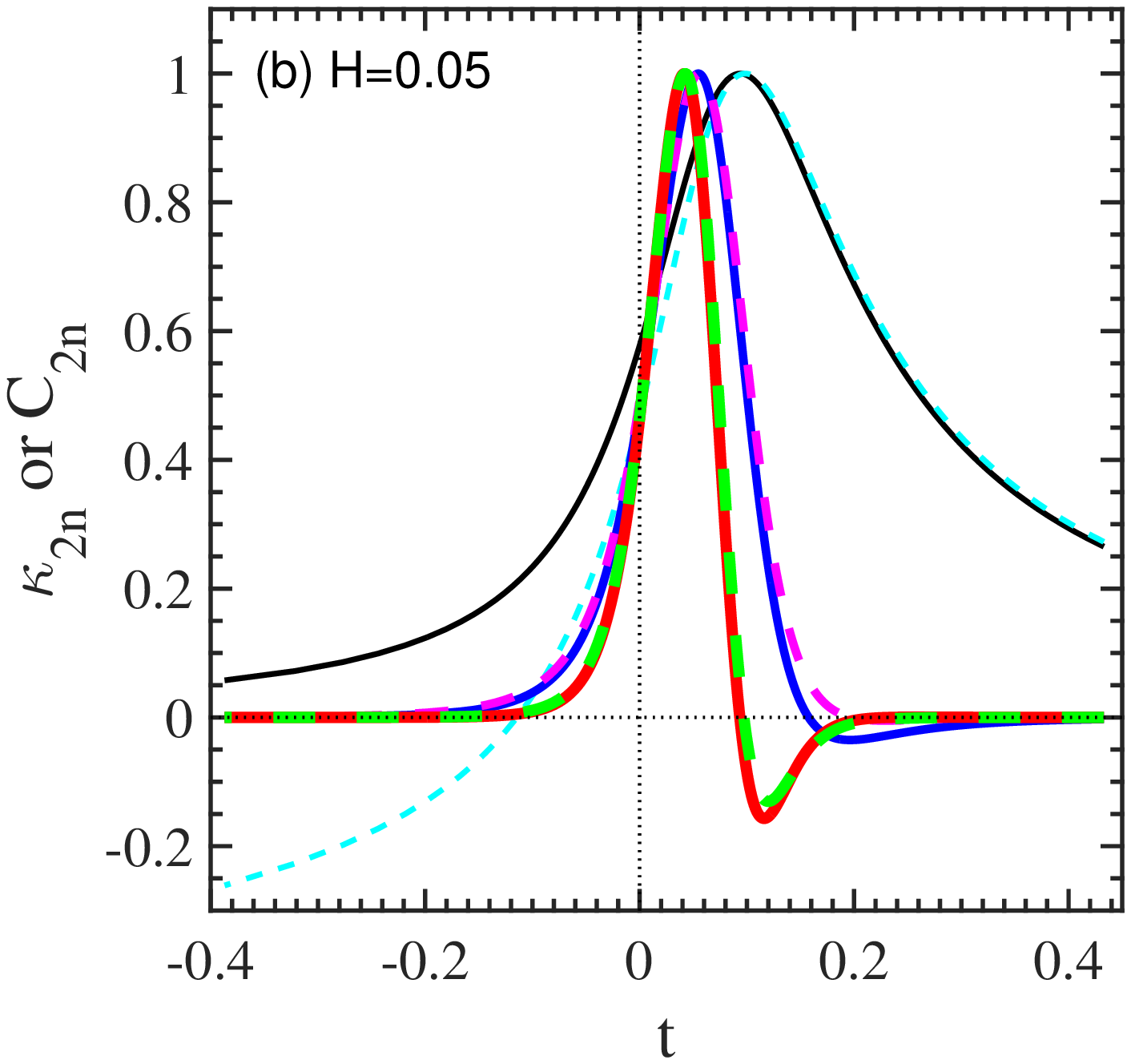}
    \includegraphics[width=0.4\textwidth]{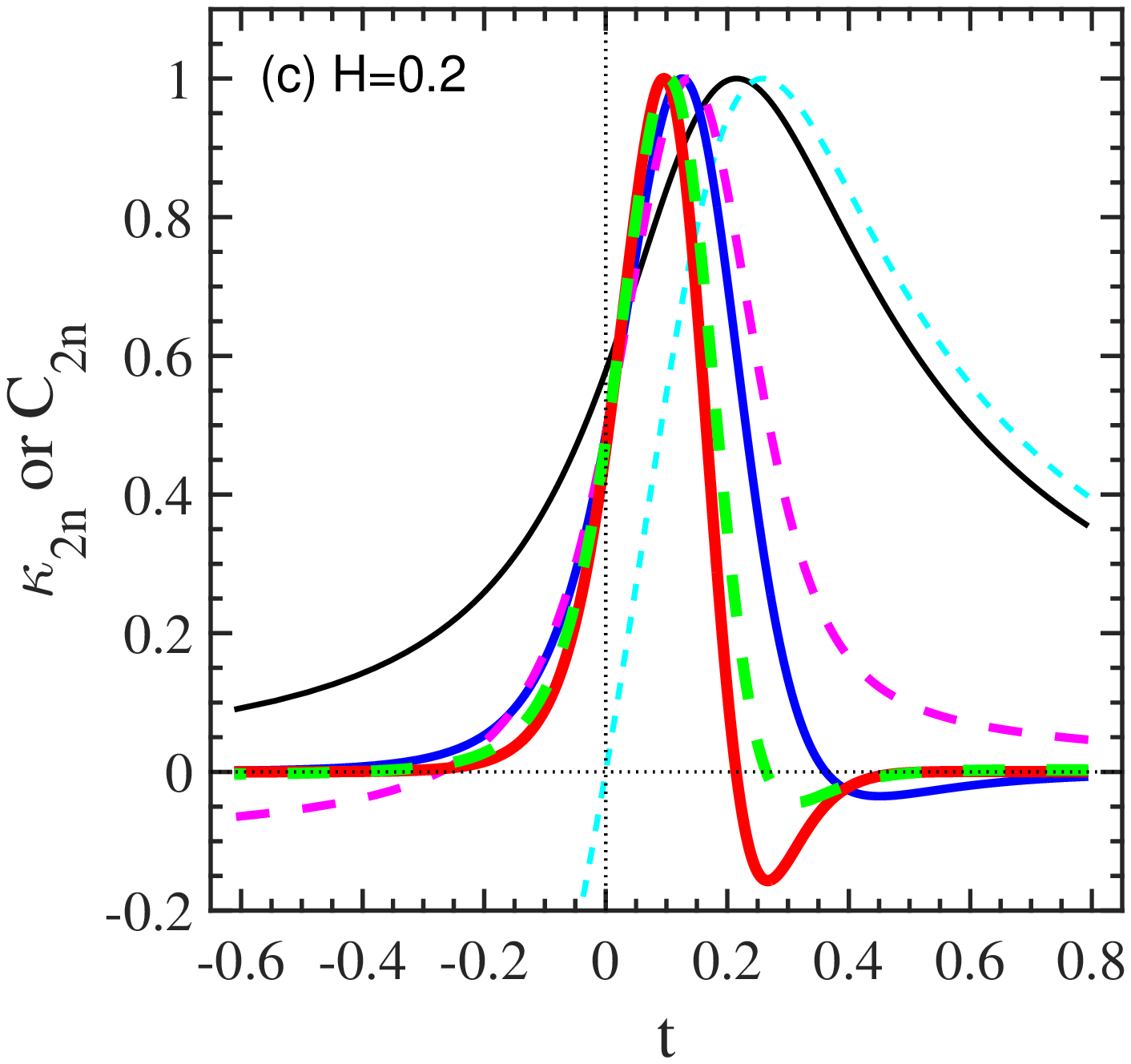}
    \includegraphics[width=0.4\textwidth]{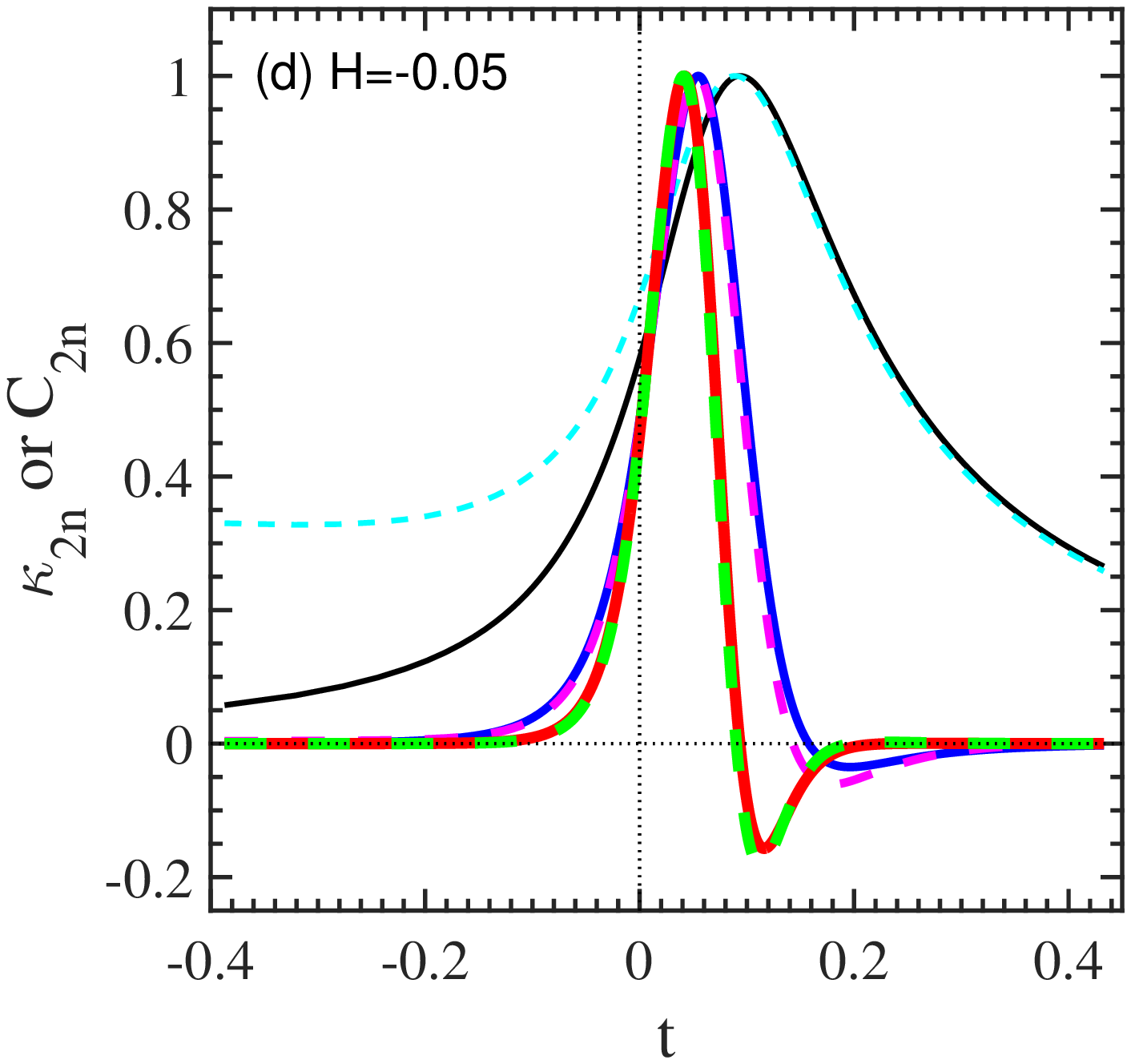}
    \caption{\label{Fig. 2}(Color online). Temperature dependence of $\kappa_{2n}$ and $C_{2n}$ for $n=1, 2, 3$ at $H=0.01$ (a), $H=0.05$ (b), $H=0.2$ (c) and $H=-0.05$ (d), respectively.}
\end{figure*}

Temperature dependence of $C_{2n}$ changes with the variation of $H$. For example, the negative values of $C_2$ and $C_4$ at the lower temperature side become more and more obvious as the increasing $H$ as shown in Fig. 2(a)-(c), the negative valley of $C_4$ at the higher temperature side changes to positive, the negative valley of $C_6$ becomes flatter. What is more, the temperature dependence of $C_{2n}$ also differs when the magnetic field turn to negative as shown in Fig. 2(b) and (d).

Comparing $\kappa_{2n}$ and $C_{2n}$, with the increasing $H$ and far away from the phase boundary, the positive peak structure in the vicinity of critical temperature does not change, is their common feature, while their differences increase in the higher or lower temperature side, even sign changes occur.

Temperature dependence of the odd-order cumulants $\kappa_{2n-1}$ and factorial cumulants $C_{2n-1}$ for $n=2, 3$ at magnetic fields $H=0.01, 0.05, 0.2, -0.05$ are shown in Fig. 3(a)-(d), respectively. For the sake of comparison, they have been normalized by their absolute values of the minimum for a positive $H$ and maximum for a negative $H$, respectively. For $\kappa_{2n-1}$, the qualitative temperature dependence does not change with magnetic field. $\kappa_3$ keeps a valley structure and negative for a positive magnetic field as shown in Fig. 3(a)-(c), and a peak structure for a negative magnetic as shown in Fig. 3(d). Temperature dependence of $\kappa_5$ oscillates. The ratios of the minimum to maximum for $\kappa_5$ approximates $-10$ at positive magnetic fields and $-0.1$ for negative magnetic fields, respectively.

\begin{figure}[hbt]
\centering
    \includegraphics[width=0.4\textwidth]{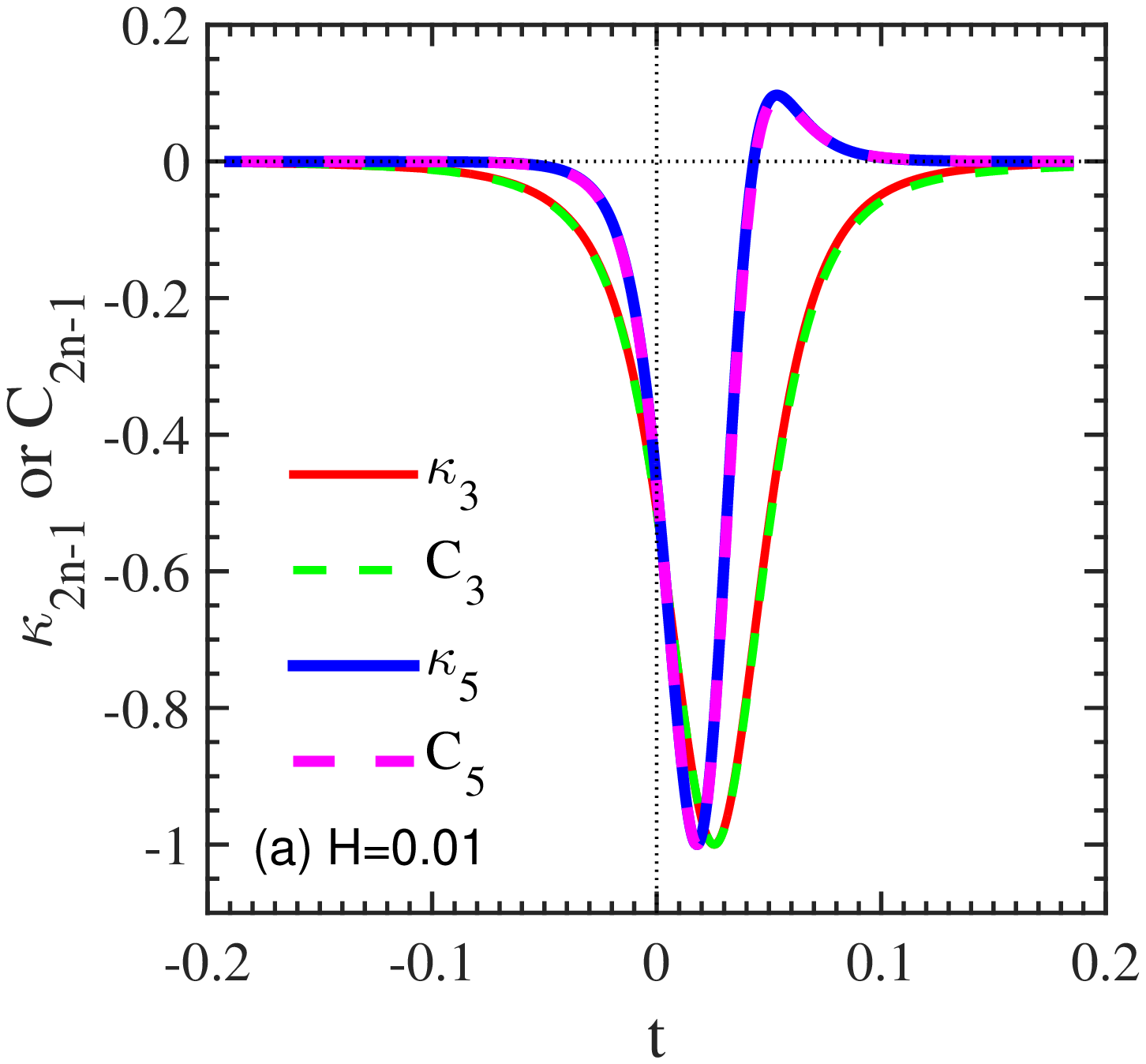}
    \includegraphics[width=0.4\textwidth]{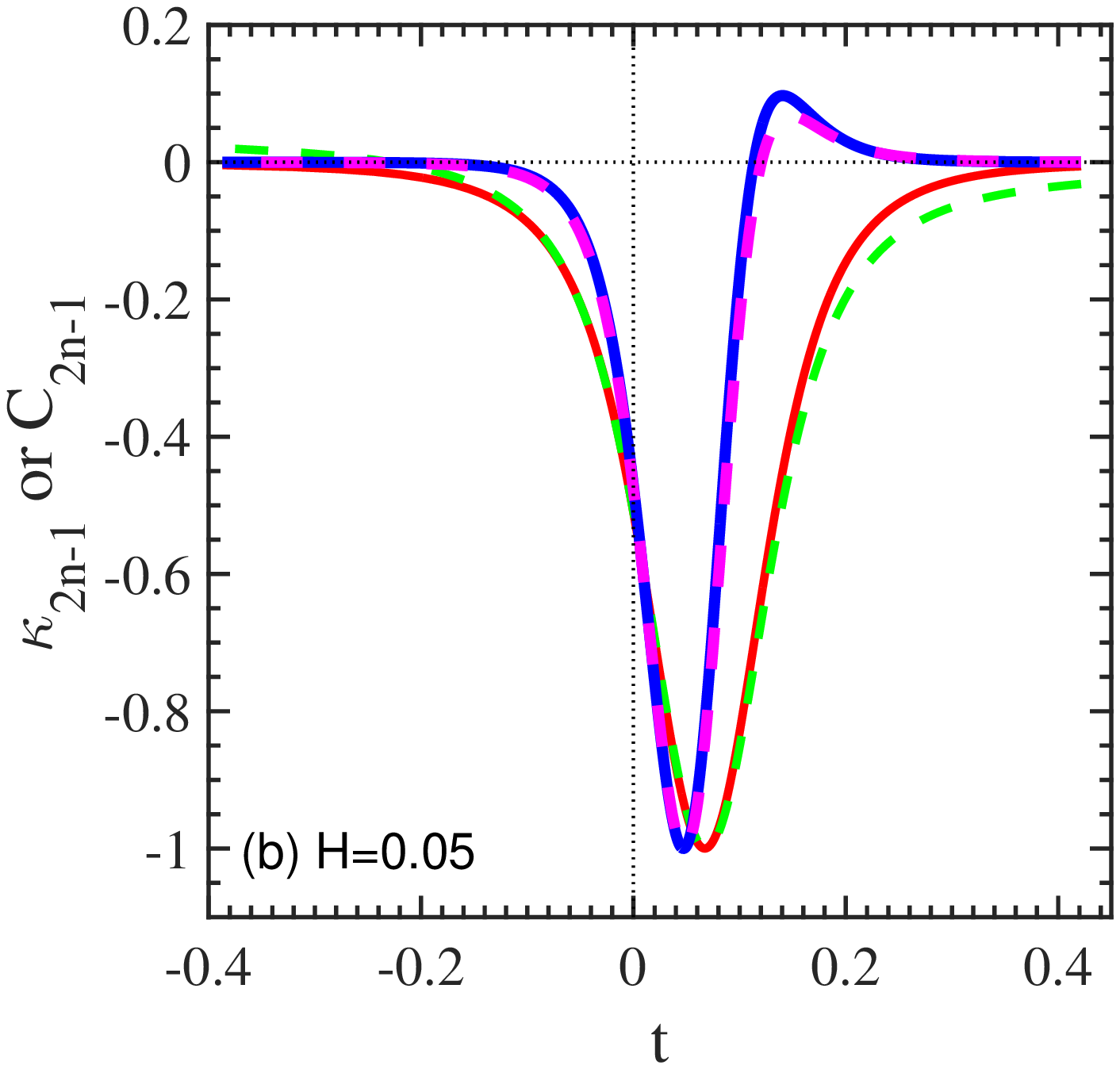}
    \includegraphics[width=0.4\textwidth]{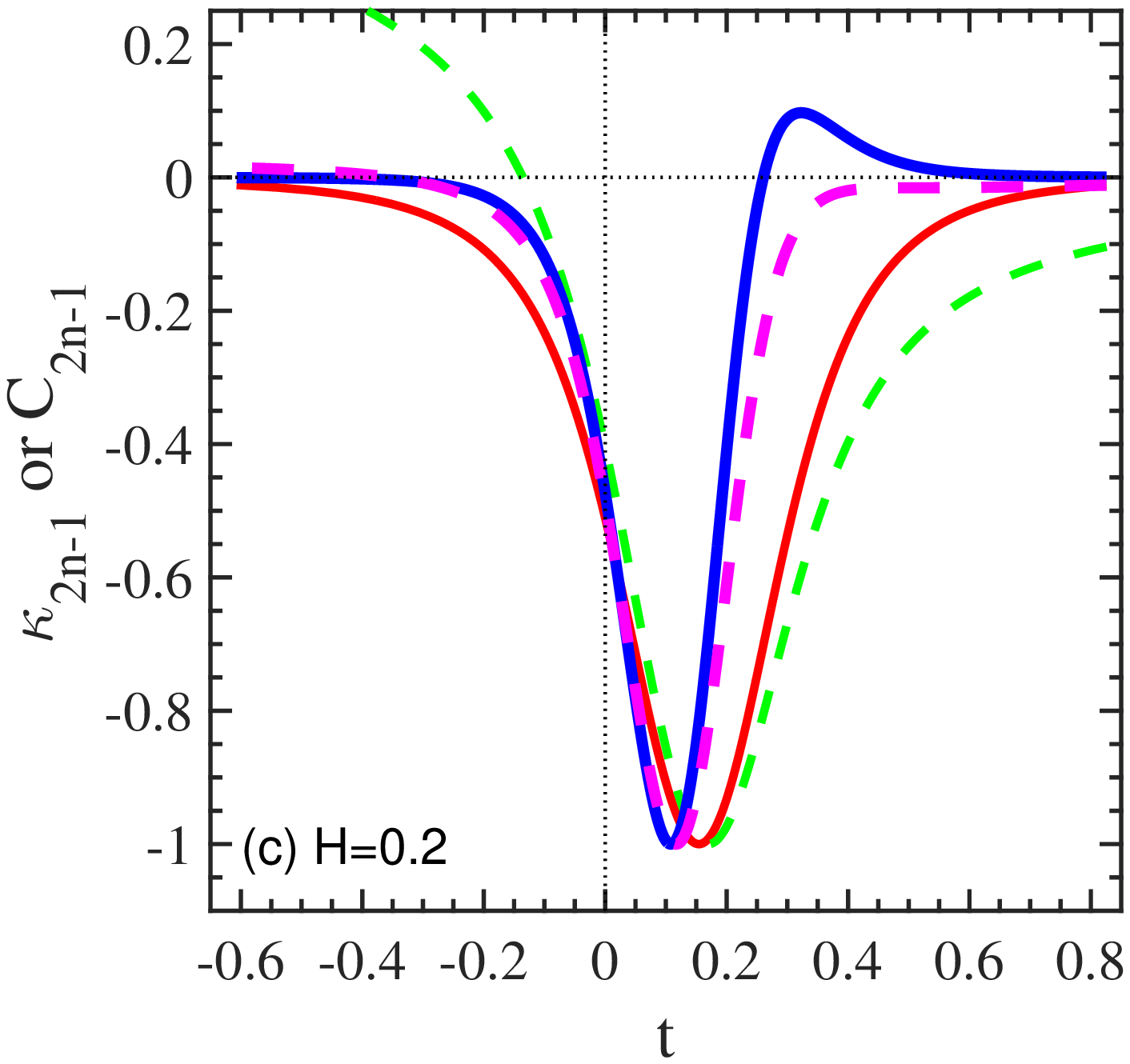}
    \includegraphics[width=0.4\textwidth]{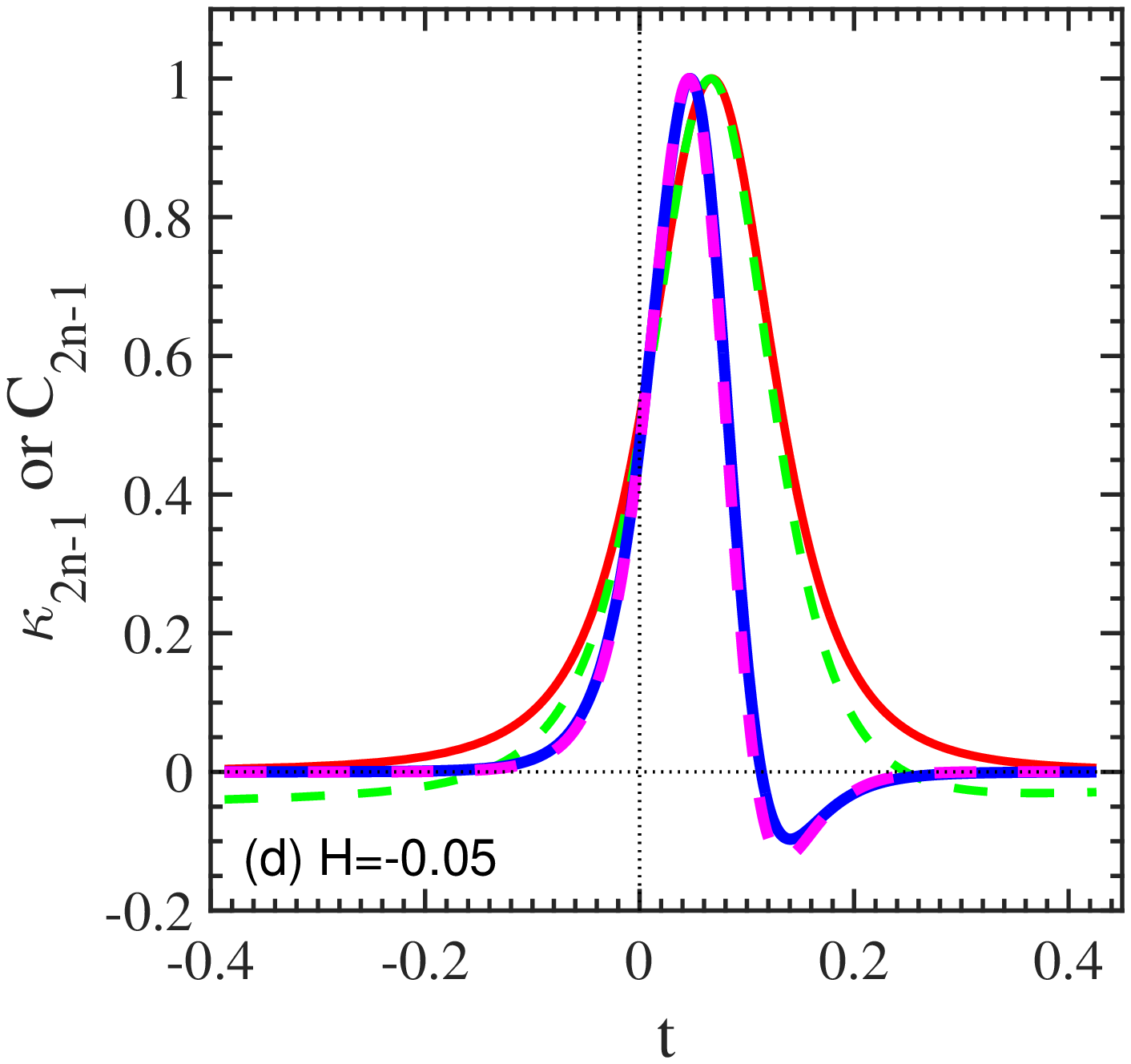}
    \caption{\label{Figure 3}(Color online). Temperature dependence of $\kappa_{2n-1}$ and $C_{2n-1}$ for $n=2, 3$ at $H=0.01$ (a), $H=0.05$ (b), $H=0.2$ (c) and $H=-0.05$ (d), respectively.}
\end{figure}

Temperature dependence of $C_{2n-1}$ changes with the variation of $H$.
It is clear that at $H=0.01$ as shown in Fig. 3(a), $\kappa_3$ and $\kappa_5$ almost overlap with $C_3$ and $C_5$, respectively. When $H$ increases and far away from the phase boundary, the negative valley structure in the vicinity of critical temperature does not change, is the common feature of $\kappa_{2n-1}$ and $C_{2n-1}$ at a positive magnetic field. For a negative magnetic field, their common feature is a
positive peak in the vicinity of the critical temperature. The difference between $\kappa_{2n-1}$ and $C_{2n-1}$ increases in the higher or lower temperature side when far away from the phase boundary, even sign changes occur.

In the vicinity of the critical point, the highest order cumulant dominates the temperature dependence of factorial cumulant, a peak or valley structure is their common feature. Cumulants and factorial cumulants can not be distinguished in the 3-dimensional Ising universality class, consistent with the model of critical fluctuations in Ref.~\cite{PhysRevC93034915}.

When it is far away from the phase boundary, at the lower or higher temperature side, the temperature dependence of factorial cumulants has great changes comparing with the corresponding cumulants, even sign changes. Comparison of the temperature dependence and sign of cumulants and factorial cumulants may help to measure the distance to the critical point.

\section{Summary and conclusions}

By using the parametric representation of the 3-dimensional Ising model, density plots of the third to sixth order cumulants and factorial cumulants have been studied and compared. We found that in the vicinity of the critical point, density plots of the same order cumulant and factorial cumulant are consistent, while far away from the critical point, signs of factorial cumulants will change under the influence of the terms related with the lower order cumulants. Especially, at the lower temperature side, signs of the factorial cumulants just depend on the term related to the first-order cumulant, i.e. the magnetization.

Through the comparison of temperature dependence of cumulants and factorial cumulants at different magnetic fields, we found that the behavior of cumulants are more stable. Its qualitative temperature dependence does not change with the variation of the distance to the phase boundary. The ratio of the peak hight and valley depth in the temperature dependence of the fourth, fifth and sixth order cumuants keep the same at different magnetic fields. While the latter changes with the distance to the phase boundary. Temperature dependence of the two kinds of cumulants is almost the same in the vicinity of the critical point, showing a positive peak or negative valley structure. While obvious sign changes of factorial cumulants occur in the higher or lower temperature side. These features may help to evaluate the distance to the critical point.

\end{document}